\documentclass[12pt,iopams,setstack,epsfig,epsf,float]{iopart}
\usepackage{cite}
\usepackage{epsf}
\usepackage{epsfig}
\usepackage{float}
\usepackage{amssymb}

\def\g{{\bf{g}}}

\def\beq{\begin{equation}}
\def\eeq{\end{equation}}
\def\beqa{\begin{eqnarray}}
\def\eeqa{\end{eqnarray}}

\def\g0{{\gamma_0}}

\eqnobysec
\begin{document}
\bibliographystyle{plain}
\input epsf

\title[Single-particle spectrum of the Anderson model]{Single-particle dynamics of the Anderson model: a two-self-energy description within the numerical renormalization group approach. }

\author{Martin R Galpin and  David E Logan}

\address{ University of Oxford, Physical and Theoretical Chemistry
Laboratory,\\ South Parks Rd, Oxford OX1~3QZ, UK}

\begin{abstract}
Single-particle dynamics of the Anderson impurity model are studied using
both the numerical renormalization group (NRG) method and the local moment
approach (LMA). It is shown that a `two-self-energy' description
of dynamics inherent to the LMA, as well as a conventional `single-self-energy' description, arise within NRG; each yielding correctly the same local 
single-particle spectrum. Explicit NRG results are obtained for the broken symmetry spectral constituents arising in a two-self-energy description, and the total
spectrum. These are also compared to analytical results obtained from the
LMA as implemented in practice. Very good agreement between the two is found, essentially on all relevant energy scales from the high-energy Hubbard satellites 
to the low-energy Kondo resonance.
\end{abstract}

%\pacs{71.27.+a Strongly correlated electron systems; heavy fermions - 
%75.20.Hr Local moment in compounds and alloys; Kondo effect, valence
%fluctuations, heavy fermions}

%\submitto{\JPCM}
%\maketitle

\section{Introduction}
\label{sec:intro}

The numerical renormalization group (NRG) method~\cite{wilson,hrk} 
forms a well established, powerful numerical technique for calculating 
the properties of quantum impurity and related models. The basic paradigm 
here is of course the celebrated  Anderson impurity model (AIM)~\cite{pwa}, 
in particular the Kondo effect arising in strong coupling  
(see~\cite{hews} for a review). NRG is not restricted to calculation of static properties, but can equally handle dynamical properties such as single-particle spectra, see e.g.~\cite{sakai,CHZ,BPH}.

Dynamics in particular pose well known difficulties for analytical 
approaches~\cite{hews}; be it in handling strong correlations in general, 
spanning the full relevant range of energy scales, satisfying the 
low-energy dictates of Fermi liquid behaviour and so on.
In recent years we have been developing a local moment approach (LMA)
to single-particle dynamics and related properties of a range of quantum
impurity models, see e.g.~\cite{LMA98,LMA02,NLD01,loga02,PAIM,PAIM03,bulla}.
Physically transparent, and technically straightforward in practice, the LMA 
is based on an underlying `two-self-energy' description.
As in Anderson's original work~\cite{pwa}, local moments are introduced explicitly
from the outset, leading to \emph{two} degenerate mean-field saddle points. 
Spin-flip tunneling between the mean-field states --- embodied in dynamical
contributions to the two associated self-energies, naturally absent at
pure mean-field level --- lifts the erstwhile spin degeneracy of the saddle
points, and restores the singlet symmetry characteristic of the local Fermi
liquid state.

As implemented in practice the LMA is of course approximate, as dictated by
the particular approximation chosen for the dynamical contributions to the two
self-energies~\cite{LMA98,LMA02,NLD01,loga02,PAIM,PAIM03}. 
But it is taken for granted that, in principle, dynamics may be obtained 
either within an underlying two-self-energy framework, or 
within the `single-self-energy' description that provides the conventional route
to single-particle dynamics. In that case, it is natural to ask whether a
two-self-energy description arises also within an
essentially exact numerical approach such as NRG. It is this issue we
consider here. We show that according to whether
even or odd RG iterations are considered, both the single- and the
two-self-energy descriptions are contained in the NRG approach; and in particular
we obtain explicit NRG results for the composite broken symmetry spectra 
inherent to a TSE description.
That in turn enables direct comparison to be made between NRG results for
the underlying spectra and those arising from the LMA as implemented in practice. 
Very good agreement is found, on all energy scales characteristic of the
problem. Results are given in \S 3, following a brief discussion of the
background theory in \S 2.

\section{Theory}
\label{sec:theor}

The AIM Hamiltonian~\cite{pwa} is given in standard notation by

\beq
\hat{H} = \sum_{\bf k, \sigma}\epsilon_{\bf k}\hat{n}_{\bf k\sigma} +
\sum_{\sigma}(\epsilon_{i} + \case{U}{2}\hat{n}_{i-\sigma})\hat{n}_{i\sigma}
+\sum_{\bf k,\sigma} V_{i\bf k}(c^{\dagger}_{i\sigma}c^{\phantom\dagger}_{\bf k\sigma}
+ h.c.)
\eeq
where the first term refers to the non-interacting host, and the second to the
impurity with local interaction $U$ and energy $\epsilon_{i}$. In strong coupling
(large $U$)
the low-energy physics of the AIM is of course that of the Kondo model~\cite{hews}.
  Our focus is the $T=0$ local impurity spectrum $D(\omega) = -\case{1}{\pi}$Im$G(\omega)$, with $G(\omega)$ ($\leftrightarrow G(t) = -i\theta(t)\langle\{c^{\phantom\dagger}_{i\sigma}(t),c^{\dagger}_{i\sigma}\}\rangle$) the impurity propagator,
\beq
G(\omega) = \frac{1}{[g^{-1}(\omega) - \Sigma(\omega)]}.
\eeq
Here $g(\omega) = [\omega^{+} -\epsilon_{i} - \Delta(\omega)]^{-1}$ is the non-interacting propagator with $\Delta(\omega) = \sum_{\bf k} V^{2}_{i\bf k}[\omega^{+} - \epsilon_{\bf k}]^{-1}$ the hybridization function, such that $\Delta(\omega) = -i\Delta_{0}$ for a wide flat-band host~\cite{hews}, with $\Delta_{0} = \pi V^{2}\rho_{\rm host}(\omega =0)$ the hybridization strength ($\omega =0$ refers to the Fermi
level, and $V \equiv V_{i\bf k}$). $\Sigma(\omega) = \Sigma^{R}(\omega)  -i \Sigma^{I}(\omega)$ denotes the interaction self-energy, which is merely defined 
by the Dyson equation implicit in equation (2.2).

  It is of course $\Sigma(\omega)$ that provides the conventional theoretical
route to dynamics, via perturbation theory (PT) in $U$~\cite{hews}.
This approach is fine in principle, order by order in PT. But it is limited 
in practice, reflecting the inability of PT to handle strong correlations in
general~\cite{hews}. The practical difficulties arise because construction of 
$\Sigma(\omega)$ via conventional PT based on Hartree-modified propagators, 
beginning with the static Hartree bubble diagram $\Sigma^{0}$, amounts 
to an expansion about the Hartree mean-field saddle point. And that 
single-determinantal saddle point is generally unstable to local moment condensation~\cite{pwa}, reflected e.g.\ in the fact that the standard diagrammatic resummations 
one might expect to be required to capture strong correlations
--- such as the sum of all particle-hole interactions in the transverse spin 
channel --- give rise to well known divergences/non-analyticities in 
$\Sigma(\omega)$~\cite{hews,LMA98,LMA02}. 

In these circumstances the LMA simply recognises~\cite{LMA98,LMA02,NLD01} 
that the natural mean-field saddle point about which to expand is 
unrestricted Hartree-Fock, corresponding to condensed local moments.
Since $\hat{H}$ is invariant under $\sigma \leftrightarrow -\sigma$, this saddle
point is now of course doubly degenerate, denoted by $\alpha = A$ or $B$
corresponding respectively to local moments $\mu = +|\mu|$ and $-|\mu|$.
Accordingly, the full $G(\omega)$ is expressed as~\cite{LMA98,LMA02,NLD01}
\numparts
\begin{eqnarray}
G(\omega) &= \case{1}{2}[G_{A\sigma}(\omega) + G_{B\sigma}(\omega)]\\
          &= \left\{\frac{1}{[g^{-1}(\omega) - \tilde{\Sigma}_{A\sigma}(\omega)]} 
+ \frac{1}{[g^{-1}(\omega) - \tilde{\Sigma}_{B\sigma}(\omega)]} \right\} 
\end{eqnarray}
\endnumparts
with self-energies $\tilde{\Sigma}_{\alpha\sigma}(\omega)$ ($ = \tilde{\Sigma}^{0}_{\alpha\sigma} + \Sigma_{\alpha\sigma}(\omega)$ with 
$\tilde{\Sigma}^{0}_{\alpha\sigma}$ the static Hartree-Fock bubble). 
Note that equation (2.3a) is rotationally invariant, since the invariance
of $\hat{H}$ under $\sigma \leftrightarrow -\sigma$ implies generally that 
$G_{A\sigma}(\omega)=G_{B-\sigma}(\omega)$ and hence that $G(\omega)$ is correctly independent of spin $\sigma$. Direct comparison of equations (2.2) and (2.3b) 
also clearly
implies a general relation (equation (3.4) of~\cite{LMA02}) between the 
single self-energy $\Sigma(\omega)$ and the 
$\{\tilde{\Sigma}_{\alpha\sigma}(\omega)\}$, enabling the former to be obtained
directly from the latter.

  At this stage we emphasise the generality of the above considerations.
The impurity propagator $G(\omega)$, and hence spectrum $D(\omega)$, may
be obtained either via the conventional single self-energy description embodied
in equation (2.2), or via the two-self-energy (TSE) description inherent in
equation (2.3): which one uses is a matter of choice, at
least in principle. In practice, of course, the stability of the underlying 
mean-field (MF) saddle point arguably renders the TSE description a more natural 
choice. 
The standard diagrammatic resummations for the dynamical contribution 
$\Sigma_{\alpha\sigma}(\omega)$ to $\tilde{\Sigma}_{\alpha\sigma}(\omega)$ no
longer suffer from non-analyticities, and  as detailed in~\cite{LMA98,LMA02,NLD01}
may be employed with impunity. This forms the practical basis of the LMA.
Moreover the relation mentioned above between $\Sigma(\omega)$ and
$\{\tilde{\Sigma}_{\alpha\sigma}(\omega)\}$ may be employed to ascertain
directly, and generally, 
the conditions on $\{\tilde{\Sigma}_{\alpha\sigma}(\omega)\}$ 
under which $\Sigma^{I}(\omega) \sim {\cal{O}}(\omega^{2})$ as $\omega \rightarrow 0$, i.e.\ exhibits Fermi liquid behaviour.
This is merely a matter of algebra, and as detailed in~\cite{LMA02} the 
requisite condition is
\beq
\tilde{\Sigma}_{A\sigma}^{R}(0) = \tilde{\Sigma}_{B\sigma}^{R}(0)
~~~~~~~~~~~~(\equiv \Sigma^{R}(0))
\eeq
referring exclusively to the Fermi level $\omega =0$. In an exact theory for
the metallic AIM,
this `symmetry restoration' condition should be satisfied automatically (see
also \S 3).
For the LMA in practice~\cite{LMA98,LMA02,NLD01} it is enforced self-consistently,
and thereby determines the local moment magnitude $|\mu|$
(supplanting the usual gap equation for $|\mu|$ at pure MF 
level)~\cite{LMA98,LMA02,NLD01}. It is
also central in being able to access the quantum phase
transition to a degenerate local moment phase where such arises, as it does in the
pseudogap AIM~\cite{WF,PAIM,PAIM03,bulla}.

How well the LMA in practice captures single-particle dynamics can of course be
tested by direct comparison to NRG calculations. For the metallic AIM, that has
been considered in~\cite{NLD01} (see also~\cite{bulla}). The resultant
LMA scaling spectrum $D(\omega)$ \emph{vs} $\omega/\omega_{K}$ in the strong coupling,
Kondo regime was shown to be in excellent agreement with NRG results.
However an issue evidently remains. Within the LMA the single-particle 
spectrum is expressed (equation (2.3)) as 
$D(\omega) = \case{1}{2}[D_{A\sigma}(\omega) +D_{B\sigma}(\omega)]$ (where
$D_{\alpha\sigma}(\omega) = -\case{1}{\pi}$Im$G_{\alpha\sigma}(\omega)$).
It is this that has hitherto been compared to NRG results. 
The considerations above suggest however that there should exist an 
NRG counterpart of the spectral densities $D_{A\sigma}(\omega)$ and 
$D_{B\sigma}(\omega)$ inherent to a TSE description.
It is this we now consider. The resultant $D_{\alpha\sigma}(\omega)$ can
in turn be individually interrogated and compared to the detailed predictions
for them arising from the LMA in practice.

\subsection{NRG spectra}

  In Wilson's NRG method~\cite{wilson,hrk} the AIM is mapped onto a 
semi-infinite chain. This is diagonalised iteratively starting from
the free impurity, and with a suitably truncated basis such that with
increasing chain length essentially only the lowest lying states are
renormalised. Spectral functions at each iteration $N$  are 
calculated~\cite{sakai,CHZ} from the appropriate matrix elements 
connecting the NRG ground state and excited states, themselves related 
recursively to those of the previous iteration; and the spectrum for
the whole frequency range is built up from the results for all iterations
(roughly speaking, iteration $N$ determines the spectrum at frequencies
$\omega \sim D\Lambda^{-(N-1)/2}$ with $D$ the conduction electron bandwidth
and $\Lambda$ the usual NRG discretisation parameter~\cite{wilson,hrk}).
The resultant spectrum consists of a set of $\delta$-functions with known
weights at discrete frequencies, which is then broadened on a logarithmic
scale to recover the continuum, see specifically equation (10) of~\cite{BPH}. 
This is the standard
approach to calculating $T=0$ spectra via the NRG, and we follow it
here. [An alternative method~\cite{BPH} focuses directly on
obtaining the self-energy $\Sigma(\omega)$. Although not used  
here, we simply remark that it can also be extended to encompass the 
TSE spectral description considered below.]

  NRG spectra may be calculated from either the even set or the odd set of
iterations (the fixed point, here exclusively strong coupling, is a fixed point
of the square of the RG transformation~\cite{wilson,hrk}). The same spectrum
$D(\omega)$ results in either case, as shown explicitly below. There is
however an important difference between even-$N$ and odd-$N$ iterations.
In the former case, the NRG ground state is always a
non-degenerate spin-singlet. For the odd-$N$ iterations by contrast we find
that the NRG ground state (for any iteration) is a degenerate doublet,
$S = \case{1}{2}$ (where we emphasise that $S$ denotes the \emph{total} 
spin angular momentum quantum number of the entire system).
Denoting the $S_{z} = +\case{1}{2}$ component of this doublet by `A', 
and the $S_{z} = -\case{1}{2}$ component by `B', we can thus construct separate
single-particle spectra (for either spin $\sigma$) from each of these 
degenerate ground states, denoted by $D_{A\sigma}(\omega)$ and $D_{B\sigma}(\omega)$
respectively; 
such that the total, normalised single-particle spectrum is given simply
by $D(\omega) = \case{1}{2}[D_{A\sigma}(\omega) + D_{B\sigma}(\omega)]$.
From the invariance of $\hat{H}$ under $\sigma \leftrightarrow -\sigma$ it follows
that $D_{A\sigma}(\omega) = D_{B-\sigma}(\omega)$; whence only (say)
$D_{A\uparrow}(\omega)$ and $D_{B\uparrow}(\omega)$ need be considered 
in general. And for the particle-hole symmetric AIM ($\epsilon_{i} = -\case{U}{2}$)
considered explicitly
below, it follows further that $D_{\alpha\sigma}(\omega) = D_{\alpha-\sigma}(-\omega)$
($\alpha = A$ or $B$), such that 
$D_{B\uparrow}(\omega) = D_{A\downarrow}(\omega) = D_{A\uparrow}(-\omega)$
and $D(\omega) = D(-\omega)$.

  The essential point of the preceding discussion is clear. NRG spectra arising
from odd-$N$ iterations generate in effect a TSE description of single-particle dynamics,
enabling both the $D_{\alpha\sigma}(\omega)$ and the total spectrum
$D(\omega)$ to be determined. The even-$N$ NRG spectrum by contrast, arising 
as it does from the non-degenerate NRG ground state in this case, corresponds
to a conventional single-self-energy description of dynamics, from
which $D(\omega)$ itself may again be obtained directly. 

\begin{figure}
\centering
{\mbox{\epsffile{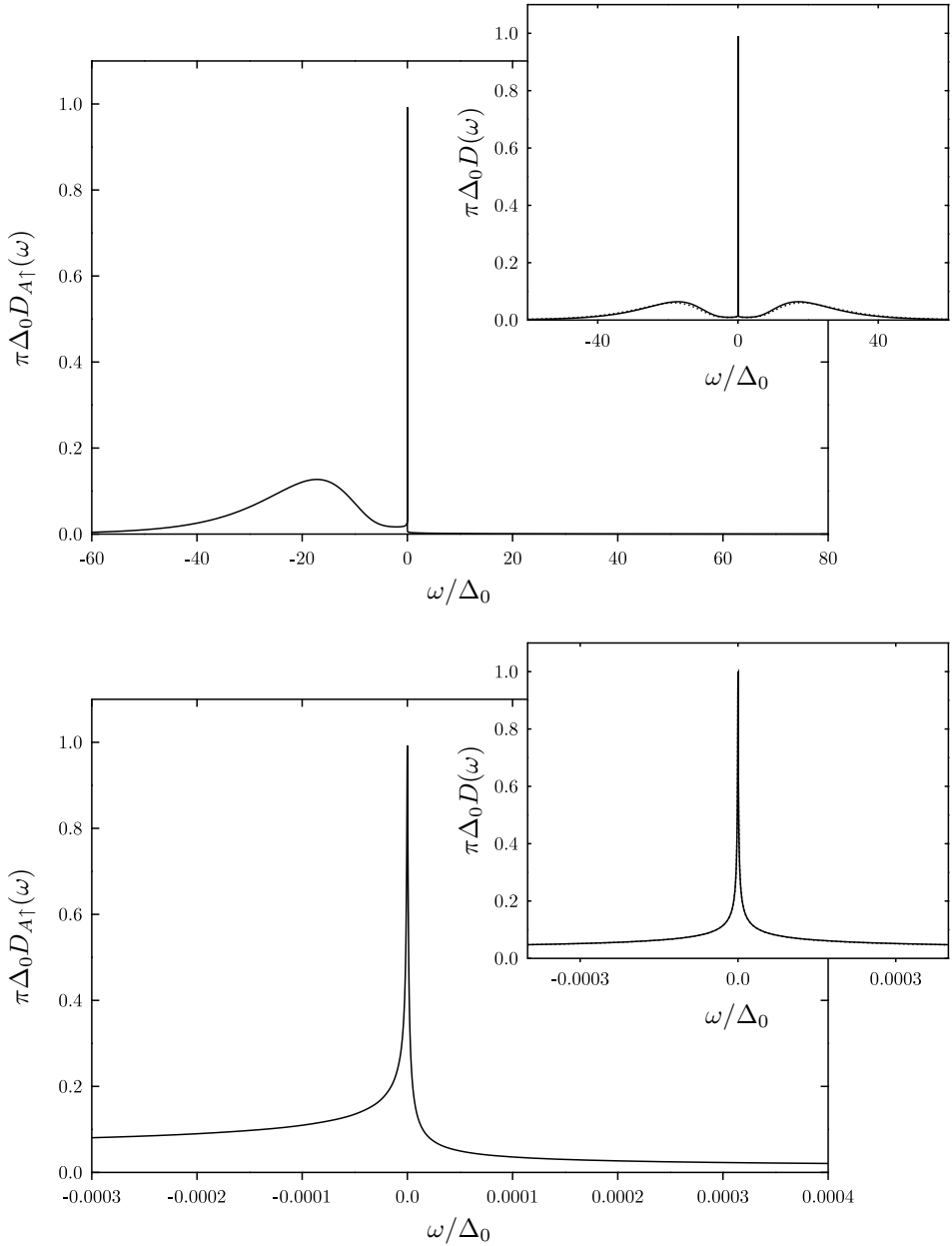}}}
\caption{NRG spectrum 
$\pi\Delta_{0}D_{A\uparrow}(\omega)$ \emph{vs} $\omega/\Delta_{0}$
for $\tilde{U} = 12$. The upper panel shows $D_{A\uparrow}(\omega)$ on 
`all scales' including
the single, high-energy Hubbard satellite and the Kondo resonance. The lower panel
focuses on the asymmetric, low-energy Kondo resonance. Insets: the corresponding total
spectrum $D(\omega) = \case{1}{2}[D_{A\uparrow}(\omega) + D_{B\uparrow}(\omega)]$
($= D(-\omega)$) obtained from the odd-$N$ iterations (solid line), \emph{and} 
$D(\omega)$ obtained from the even-$N$ iterations (dotted line);
the two are in fact indistinguishable.
}
\label{fig1}
\end{figure}

\section{Results: NRG and LMA}

  To illustrate the above, figure \ref{fig1} shows the resultant odd-$N$ NRG spectrum 
$D_{A\uparrow}(\omega)$ \emph{vs} $\omega/\Delta_{0}$, for the symmetric AIM
with a strong coupling interaction strength $\tilde{U} = U/\pi\Delta_{0} = 12$ (calculated with 
$\Lambda =2$, retaining $\sim 10,000$ states per iteration). We consider first
the insets, which show the corresponding total spectrum  
$D(\omega) = \case{1}{2}[D_{A\uparrow}(\omega) + D_{B\uparrow}(\omega)]$ 
from the odd-$N$ iterations (solid line), as well as
$D(\omega)$ obtained from the even-$N$ iterations (dotted line). The resultant
symmetric spectra consist as expected of two high-energy Hubbard satellites, 
symmetrically disposed at $\omega \sim \pm\case{U}{2}$, and the low-energy Kondo 
resonance.The first fact to note is that the $D(\omega)$ obtained from
odd and even iterations are indistinguishable. This underscores the point made
in \S 2: in an essentially exact approach, such as NRG, the two-self-energy
and single-self-energy descriptions are fundamentally equivalent, and which one
employs is a matter of choice. We also add (\emph{cf} \S 2) that the symmetry restoration condition equation (2.4), 
$\tilde{\Sigma}_{A\sigma}^{R}(0) = \tilde{\Sigma}_{B\sigma}^{R}(0)$ ($ =0$ 
for the symmetric AIM) is as expected satisfied within NRG, reflected
in the fact that $\pi\Delta_{0}D_{A\uparrow}(\omega =0) =
\pi\Delta_{0}D_{A\downarrow}(\omega =0) =1$ (as required by the Friedel sum rule/Luttinger integral theorem~\cite{hews}, and satisfied in practice to 
$\lesssim 1\%$ accuracy in the present calculations).

  Turning now to the main panels in figure \ref{fig1}, the obvious feature is the
intrinsic asymmetry in $D_{A\uparrow}(\omega)$. Only a lower Hubbard satellite
is seen in $D_{A\uparrow}(\omega)$ (the upper satellite correspondingly
arises in $D_{B\uparrow}(\omega) = D_{A\uparrow}(-\omega)$); consistent
with the expectation that the `A' state ($S_{z} = +\case{1}{2}$) 
connects in the atomic limit ($V_{i\bf k} =0$) to a purely 
$\uparrow$-spin occupied impurity, from which an $\uparrow$-spin electron
may be removed but not added.
The asymmetry in $D_{A\uparrow}(\omega)$ is not moreover confined to the high-energy
spectrum highlighted in the upper panel, which has an obvious counterpart
in the atomic limit: the low-energy Kondo resonance
(lower main panel) is likewise seen to be strongly asymmetric.

\begin{figure}
\centering
{\mbox{\epsffile{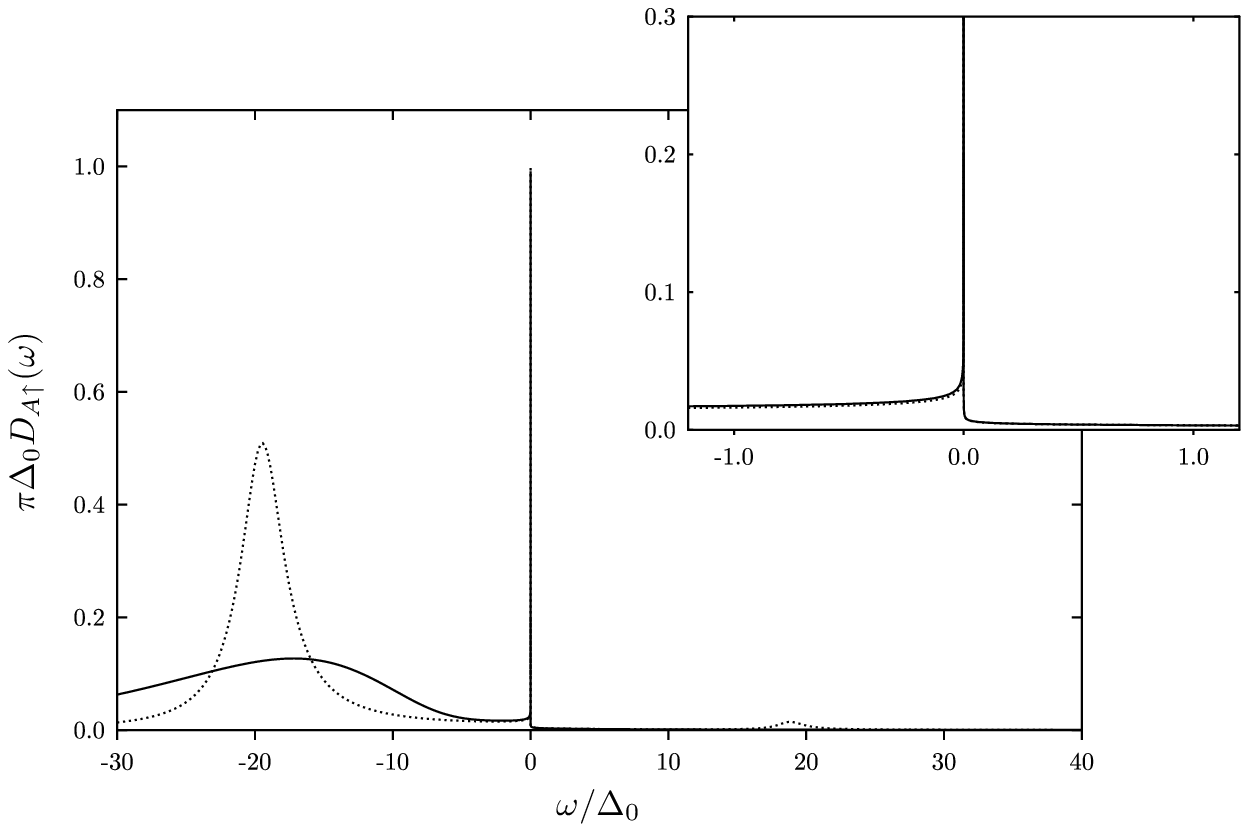}}}
\caption{$\pi\Delta_{0}D_{A\uparrow}(\omega)$ \emph{vs} $\omega/\Delta_{0}$ for
$\tilde{U}=12$, obtained from NRG (solid line) and LMA (dotted line). The inset
continues the comparison on a lower energy scale. NRG is known to overbroaden
the high energy Hubbard satellite; this is reconstructed in figure 6
below, where further comparison is made between NRG and LMA.
}
\label{fig2}
\end{figure}
  To gain further insight into these results we now consider direct comparison with
the LMA in practice, which affords an analytical handle on the single-particle
dynamics. Full details of the underlying calculations are given in~\cite{LMA98}
for the symmetric AIM considered here (and in~\cite{LMA02} for the asymmetric
case). Here we focus purely on results arising. 
  
  In figure \ref{fig2} NRG and LMA results for
$\pi\Delta_{0}D_{A\uparrow}(\omega)$ are compared,  \emph{vs} $\omega/\Delta_{0}$ 
(i.e.\ on an `absolute' energy scale). The chosen $\tilde{U} =12$ is simply
representative of the large-$\tilde{U}$, strong coupling behaviour of interest; similar results to those that follow have naturally been obtained for a range of interaction strengths. Let us first comment on the lower Hubbard satellite in 
$D_{A\uparrow}(\omega)$ (main panel), which clearly contains the vast majority of the spectral weight. The LMA satellite is itself a Lorentzian, with a HWHM of
$2\Delta_{0}$  --- twice that occurring in the pure mean-field limit (for
the physical reasons explained in~\cite{LMA98}), and
which behaviour is believed to be asymptotically exact in strong coupling.
With NRG by contrast the Hubbard satellite is well known to be 
overbroadened, due to the associated broadening on a logarithmic scale~\cite{sakai,CHZ} (as occurs also, although ameliorated somewhat, with the method introduced 
in~\cite{BPH}). We will return to this below (see figure 6), and show 
that the NRG results for the satellite are in fact consistent with 
the above Lorentzian behaviour.

\begin{figure}
\centering
{\mbox{\epsffile{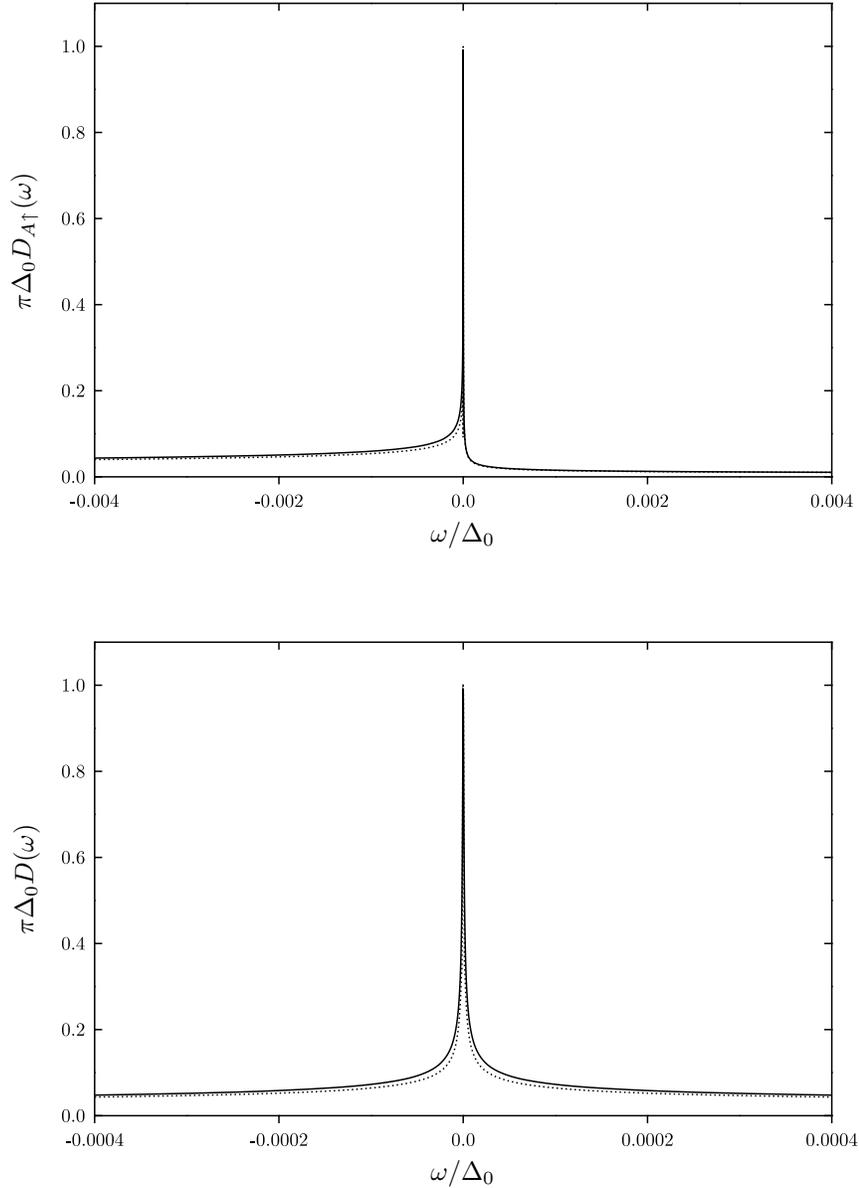}}}
\caption{For $\tilde{U} =12$, comparison between NRG (solid lines) and LMA
(dotted lines) on the lowest energy scales appropriate to the Kondo resonance.
Upper panel: $\pi\Delta_{0}D_{A\uparrow}(\omega)$ \emph{vs} $\omega/\Delta_{0}$.
Lower panel: the full spectrum $\pi\Delta_{0}D(\omega)$ \emph{vs} $\omega/\Delta_{0}$,
with $D(\omega) = \case{1}{2}[D_{A\uparrow}(\omega) + D_{B\uparrow}(\omega)]$.
}
\label{fig3}
\end{figure}

 Aside from the high energy satellite issue, NRG and LMA results 
are seen from figure \ref{fig2} to be in remarkably good agreement. That this persists down to much lower energy scales is shown further in figure \ref{fig3},
where $D_{A\uparrow}(\omega)$ is shown in the top panel and the full 
symmetric spectrum $D(\omega)$ in the lower panel. Our aim now is to obtain a 
handle on the characteristic asymmetry evident in the $D_{A\uparrow}(\omega)$ Kondo resonance. This may be obtained analytically from the LMA~\cite{NLD01}, formally
for $|\tilde{\omega}| \gg 1$ where $\tilde{\omega} = \omega/\omega_{K}$ and
the Kondo scale $\omega_{K}$ is (here defined as) the HWHM of the Kondo resonance
in $D(\omega)$. On the positive frequency side, for 
$\tilde{\omega} = |\tilde{\omega}| \gg 1$, the LMA gives the asymptotic 
behaviour~\cite{NLD01}
\beq
\pi\Delta_{0}D_{A\uparrow}(\omega) = \frac{1}{[\frac{4}{\pi}\ln(a|\tilde{\omega}|)]^{2} + 1}
\eeq
with $a \simeq 0.7$ a pure constant; while for negative frequencies by contrast, 
and $-\tilde{\omega} = |\tilde{\omega}| \gg 1$, 
\beq
\pi\Delta_{0}D_{A\uparrow}(\omega) = \frac{5}{[\frac{4}{\pi}\ln(a|\tilde{\omega}|)]^{2} + 25}.
\eeq
\begin{figure}
\centering
{\mbox{\epsffile{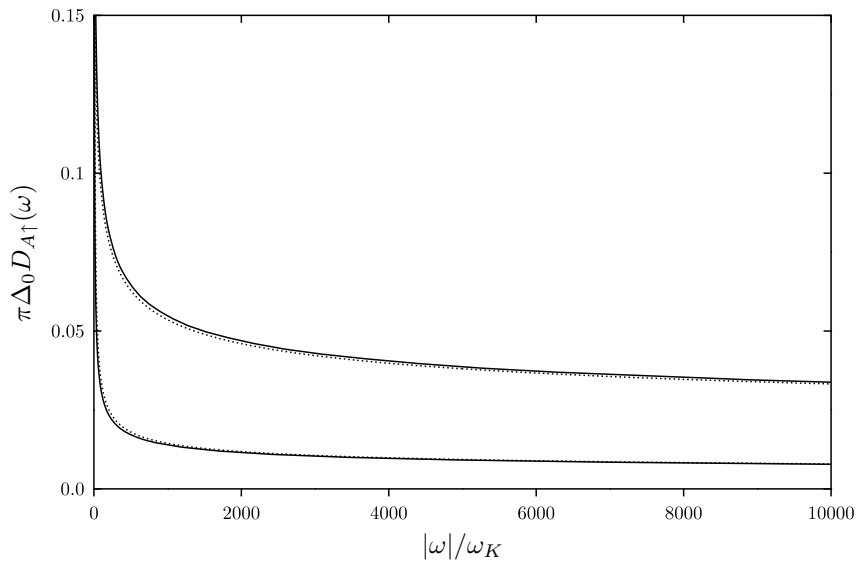}}}
\caption{Comparison between NRG results for $\pi\Delta_{0}D_{A\uparrow}(\omega)$
\emph{vs} $|\omega|/\omega_{K}$ (solid lines) and the predicted asymptotic behaviour equations  (3.1,2) (dotted lines). Lower curves: for positive frequencies, 
$\omega >0$; upper curves: for $\omega <0$.
}
\label{fig4}
\end{figure}

  Figure \ref{fig4} compares NRG results for $\pi\Delta_{0}D_{A\uparrow}(\omega)$ with this predicted asymptotic behaviour. The agreement is seen to be very good, and in practice equations (3.1,2) already describe the spectra quite
accurately for $|\omega|/\omega_{K} \gtrsim 2$ or so. We also add that the full
forms of equations (3.1,2) are required for the agreement shown in figure \ref{fig4}, i.e.\ the
behaviour in the $|\omega|/\omega_{K}$-range shown is not dominated by the ultimate
high-frequency asymptotic behaviours of $\pi^{2}/[16\ln^{2}(|\omega|/\omega_{K})]$
and $5\pi^{2}/[16\ln^{2}(|\omega|/\omega_{K})]$. For the full spectrum
$D(\omega) = \case{1}{2}[D_{A\uparrow}(\omega) + D_{B\uparrow}(\omega)]$ $\equiv
\case{1}{2}[D_{A\uparrow}(\omega) + D_{A\uparrow}(-\omega)]$ the corresponding
result is obviously just the weighted sum of equations (3.1,2) (with ultimate asymptotic behaviour of $3\pi^{2}/[16\ln^{2}(|\omega|/\omega_{K})]$ that is
exact for the $s=\case{1}{2}$ Kondo model~\cite{NLD01}). It has been shown in~\cite{NLD01} that this result for $D(\omega)$ itself is in excellent agreement with NRG
results. We regard it as remarkable that the LMA provides an equally 
compelling description of the asymmetric $D_{\alpha\sigma}(\omega)$.

\begin{figure}
\centering
{\mbox{\epsffile{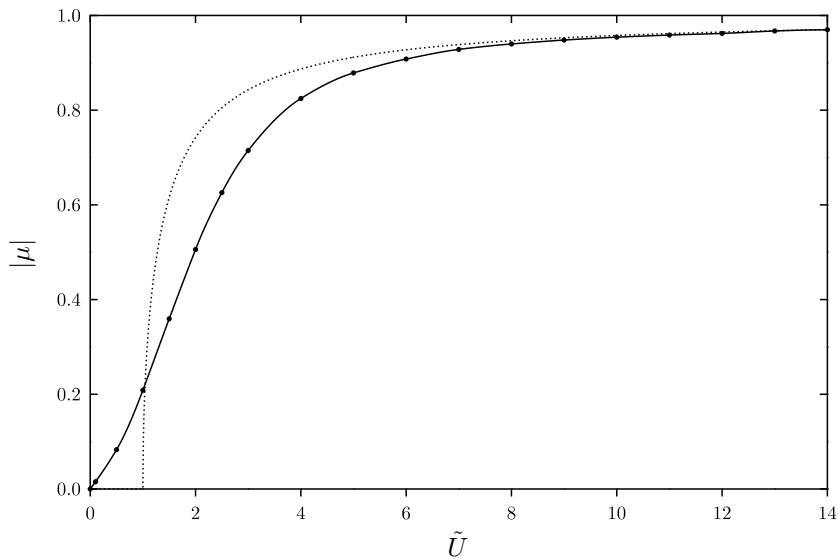}}}
\caption{NRG local moment 
$|\mu| = \int^{0}_{-\infty}d\omega ~ [D_{A\uparrow}(\omega) - D_{A\downarrow}(\omega)]$
\emph{vs} $\tilde{U} = U/\pi\Delta_{0}$ (solid line); compared to its mean-field
counterpart (dotted line) which in strong coupling $\tilde{U} \gg 1$ gives
$|\mu| \sim 1 - 4/\pi^{2}\tilde{U}$. The latter result is in fact
asymptotically exact, and is seen to be recovered by NRG.
}
\label{fig5}
\end{figure}

  The fact that the $D_{\alpha\sigma}(\omega)$ are strongly asymmetric also means of
course that the degenerate odd-$N$ NRG ground states `A' and `B' carry a local
moment, equal and opposite for the two states, with magnitude given by
$|\mu| = \int^{0}_{-\infty}d\omega ~ [D_{A\uparrow}(\omega) - D_{A\downarrow}(\omega)]$.
The resultant
$|\mu|$ is shown in figure \ref{fig5} as a function of $\tilde{U}$, and compared to its pure mean-field counterpart. The latter is given from solution of 
$|\mu| = \case{2}{\pi}\tan^{-1}(\case{U|\mu|}{2\Delta_{0}})$~\cite{footnote} (and in 
practice the LMA $|\mu|$ determined from symmetry restoration is very close to
the MF value, exponentially so in strong coupling $\tilde{U} \gg 1$~\cite{LMA98}).
Recognising that in strong coupling the Kondo resonance carries exponentially small weight and thus makes essentially no contribution to the resultant $|\mu|$, the 
leading asymptotic behaviour of the moment may be determined exactly from second
order perturbation theory in the hybridization $V_{i\bf k}$. It is found thereby 
to be given by $|\mu| \sim 1 - 4/\pi^{2}\tilde{U}$, which amusingly is precisely 
the leading result obtained from pure mean-field, and which as seen from figure 
\ref{fig5} is indeed recovered correctly from the NRG calculations.

\begin{figure}
\centering
{\mbox{\epsffile{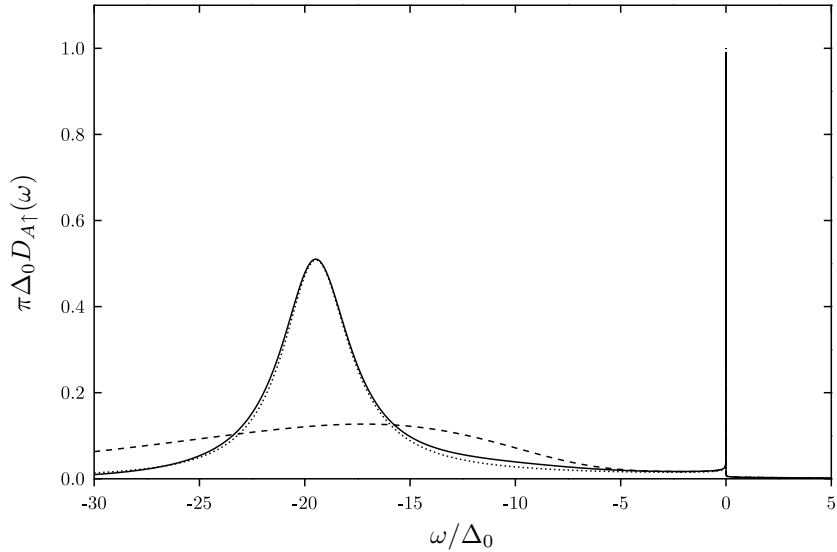}}}
\caption{$\pi\Delta_{0}D_{A\uparrow}(\omega)$ \emph{vs} $\omega/\Delta_{0}$
for $\tilde{U} =12$ obtained from the LMA (dotted line) and NRG (solid line),
with the NRG Hubbard satellite obtained as described in the text. The agreement
is now excellent on all energy scales. The original, overbroadened NRG spectrum
is shown for comparison (dashed line).
}
\label{fig6}
\end{figure}

  We return finally to the issue of the overbroadened NRG Hubbard satellite, 
and show that the `raw' NRG results are entirely consistent with a Lorentzian satellite
of width $2\Delta_{0}$. To that end we calculate, as a sum of poleweights, the cumulative NRG spectrum $F_{\rm NRG}(\omega) = \int^{\omega}_{-\infty}d\omega~
D_{A\uparrow}(\omega)$; the corresponding result for the pure Lorentzian would be 
$F_{L} = \case{1}{\pi}\tan^{-1}(\case{\omega -\omega_{0}}{2\Delta_{0}}) + \case{1}{2}$
with $\omega_{0} \sim -\case{U}{2}$ the satellite maximum. 
Writing $F_{\rm NRG}(\omega) = F_{L}(\omega) + \delta F(\omega)$ with 
$\delta F(\omega)$ thus defined, the NRG poles in $D_{A\uparrow}(\omega)$
contributing to $\delta F(\omega)$ are broadened in the
usual logarithmic fashion, and added to the pure Lorentzian contribution (the net spectral weight below the Fermi level is of course preserved). If the NRG results
are consistent with the Lorentzian, little deviation from this form should result.
That this is indeed the case is shown in figure \ref{fig6}, and NRG and LMA results 
now agree well on all energy scales including the high-energy Hubbard satellite.

\section{Concluding remarks}

  In this paper we have considered first a general issue: does a
two-self-energy description of dynamics that is inherent to the local moment approach
arise also within the numerical renormalization group method?
The answer to that question is clearly yes --- both it and a conventional
single-self-energy framework arise naturally within NRG, according to
whether odd or even RG iterations are considered. In consequence, explicit NRG 
results for the composite broken symmetry spectra underlying the two-self-energy description may be obtained:
$D_{A\sigma}(\omega)$ and $D_{B\sigma}(\omega)$
such that $D(\omega) =\case{1}{2}[D_{A\sigma}(\omega) + D_{B\sigma}(\omega)]$
gives the total impurity spectrum (and with $D(\omega)$ coincident for both
odd/even iterations). 
These in turn have been compared to results arising from
the LMA as it is implemented in practice, and very good agreement found on 
essentially all characteristic energy scales from the high-energy Hubbard 
satellites to the low-energy Kondo resonance.

  We also add that our essential conclusion is naturally not specific 
to the metallic Anderson impurity model considered here: a two-self-energy
description will arise for essentially any quantum impurity model, such as
the pseudogap AIM~\cite{WF,PAIM,PAIM03,bulla}
(where the TSE description is in general a necessity and not a 
luxury), and more generally for lattice-based models such as the periodic Anderson
model within the framework of dynamical mean-field theory~\cite{PAM}.

\ack
We are grateful to the EPSRC for supporting this research.

%\end{document}


\begin{thebibliography}{100}
\bibitem{footnote} We consider explicitly the usual `wide band' case where the 
conduction bandwidth $D$ is by far the largest energy scale.
\end{thebibliography}
\end{document}